\documentclass[twocolumn,english,pra]{revtex4}
\usepackage[T1]{fontenc}
\usepackage[latin9]{inputenc}
\usepackage{amsmath}
\usepackage{amssymb}
\usepackage{graphicx}
\usepackage{esint}

\makeatletter
\@ifundefined{textcolor}{}
{%
 \definecolor{BLACK}{gray}{0}
 \definecolor{WHITE}{gray}{1}
 \definecolor{RED}{rgb}{1,0,0}
 \definecolor{GREEN}{rgb}{0,1,0}
 \definecolor{BLUE}{rgb}{0,0,1}
 \definecolor{CYAN}{cmyk}{1,0,0,0}
 \definecolor{MAGENTA}{cmyk}{0,1,0,0}
 \definecolor{YELLOW}{cmyk}{0,0,1,0}
}

\makeatother

\usepackage{babel}
\begin{document}

\title{Virial expansion of a harmonically trapped Fermi gas across a narrow
Feshbach resonance}

\author{Shi-Guo Peng$^{1}$}

\author{Shuo-Han Zhao$^{1}$}

\author{Kaijun Jiang$^{1,2}$}

\email{kjjiang@wipm.ac.cn}

\selectlanguage{english}%

\affiliation{$^{1}$State Key Laboratory of Magnetic Resonance and Atomic and
Molecular Physics, Wuhan Institute of Physics and Mathematics, Chinese
Academy of Sciences, Wuhan 430071, China}

\affiliation{$^{2}$Center for Cold Atom Physics, Chinese Academy of Sciences,
Wuhan 430071, China}

\date{\today}
\begin{abstract}
We theoretically investigate the high-temperature thermodynamics of
a\emph{ harmonically trapped} Fermi gas across a narrow Feshbach resonance,
by using the second-order quantum virial expansion, and point out
some new features compared to the broad resonance. The interatomic
interaction is modeled by the pseudopotential with an additional parameter,
i.e., the effective range, to characterize the narrow resonance width.
Deeply inside the width of a narrow Feshbach resonance, we find the
second virial coefficient evolves with the effective range from the
well-known universal value $1/4$ in the broad-resonance limit to
one another value $1/2$ in the narrow-resonance limit. This means
the Fermi gas is more strongly interacted at the narrow resonance.
In addition, far beyond the resonance width, we find the harmonically
trapped Fermi gas still manifests appreciable interaction effect across
a narrow Feshbach resonance, which is contrary to our knowledge of
the broad Feshbach resonance. All our results can be directly tested
in current narrow Feshbach resonance experiments, which are generally
carried out in a harmonic trap.
\end{abstract}


\maketitle

\section{introduction}

Recently, the accurate measurements of $^{6}$Li atoms across the
narrow Feshbach resonance at $B_{0}=534.3\text{G}$ with the width
$\Delta=0.1\text{G}$ \cite{Hazlett2012R} has raised a great deal
of interest on narrow resonances. Ho\emph{ et al.} have shown that
for narrow resonances, the system is more strongly interacted, and
the interaction energy is highly asymmetric \cite{Ho2012A}. Cui finds
very different thermodynamic properties between broad and narrow resonances
in quasi-one-dimensional geometry \cite{Cui2012Q}. In addition, Nishida
points out that a three-component Fermi gas near a narrow Feshbach
resonance should have a universal ground state, which is absent for
a broad resonance due to the Thomas collapse \cite{Nishida2012N}.
Therefore, the accurate manipulation of ultracold atoms across a narrow
Feshbach resonance provides another experimental platform for studying
fundamental problems in condensed matters and the strongly correlated
many-body physics.

Quantum virial expansion method is a natural bridge between few-body
and many-body physics \cite{Rupak2007U,Liu2009V,Peng2011H,Liu2013V},
and the latter is always a challenge \cite{Ho2004U,Hu2007U}. The
biggest advantage of the virial expansion is that the thermodynamics
of a profound many-body system may be evaluated perturbatively from
the simple few-body solutions. For broad resonances, the second virial
coefficient for a homogeneous two-component Fermi gas has been calculated
\cite{Ho2004H}, while the second and third virial coefficients of
a strongly interacting Fermi gas in a harmonic trap were also obtained
\cite{Liu2009V,Liu2013V}. Recently, these theoretical results were
already confirmed experimentally \cite{Nascimbene2010E}. 

However, for narrow Feshbach resonances, one additional parameter,
namely the effective range, should be required to characterize the
energy-dependent interaction between atoms, besides the scattering
length \cite{Werner2008V,Gurarie2007R,Peng2012T}. This energy-dependent
interaction may apparently affect the thermodynamic universality of
the system, which should be quite different from that of the broad-resonance
system \cite{Ho2012A}. All the previous works only considered the
homogeneous systems across the narrow resonance \cite{Ho2012A,Cui2012Q},
while the real experiments are generally carried out in a harmonic
trap. In this paper, we aim to investigate how the resonance width,
or the effective range, changes the high-temperature thermodynamics
of a \emph{harmonically trapped} Fermi gas, by using the second-order
virial expansion. The second virial coefficient is obtained from the
two-body solution, in which the interatomic interaction is modeled
by the pseudopotential with an energy-dependent scattering length
\cite{Bolda2002E}, or including an additional effective range \cite{Peng2011H,Peng2011N}.
We find, deeply inside the resonance width, the second virial coefficient
continuously evolves from the well-known universal value $1/4$ in
the broad-resonance limit to one another value $1/2$ in the narrow-resonance
limit. This means the Fermi gas is more strongly interacted at the
narrow resonance, compared to the situation of the broad resonance.
Interestingly, far beyond the resonance width, the harmonically trapped
Fermi gas still manifests appreciable interaction effect across a
narrow Feshbach resonance, which is contrary to our knowledge of the
broad Feshbach resonance. All our results can be directly tested in
current experimental platform of narrow Feshbach resonances.

This paper is arranged as follows. We firstly summarize the crucial
calculation methods of the virial expansion used in this paper in
Sec.\ref{sec:virial-expansion-method}. Subsequently, in Sec.\ref{sec:thermodynamics-deeply-inside},
the second virial coefficient for a harmonically trapped Fermi gas
is calculated deeply inside the narrow-resonance width. Based on this,
the thermodynamics of the system is also discussed in detail. In Sec.\ref{sec:thermodynamics-far-beyond},
we investigate the thermodynamics of a harmonically trapped Fermi
gas far beyond the width of a narrow Feshbach resonance. Then we present
a simple toy model in Sec.\ref{sec:about-narrow-resonances}, which
is helpful for understanding the origin of the energy-dependent interaction.
Finally, our main results are summarized in Sec.\ref{sec:conclusions}.

\section{virial expansion method\label{sec:virial-expansion-method}}

Virial expansion is a perturbation method to explore the high-temperature
thermodynamics of strongly interacting Fermi gases based on the knowledge
of few-body solutions. In this section, we will summarize the crucial
calculation methods of virial expansion used in this paper, although
virial expansion has been studied in depth \cite{Liu2009V,Liu2013V}.

The basic idea of virial expansion is that at high temperature all
the thermodynamic quantities can be expanded in series of the fugacity
$z\equiv e^{\mu/\left(k_{B}T\right)}$ , which is small since the
chemical potential $\mu$ is large and negtive at high temperature.
Here, $k_{B}$ is the Boltzmann constant. For example, the thermodynamic
potential may be written as
\begin{equation}
\Omega=-k_{B}TQ_{1}\left(z+b_{2}z^{2}+\cdots+b_{n}z^{n}+\cdots\right),\label{eq:3.1}
\end{equation}
 where $Q_{n}\equiv\text{Tr}\left[e^{-\mathcal{H}_{n}/\left(k_{B}T\right)}\right]$
is the $n$-particle partition function, and $\mathcal{H}_{n}$ is
the total Hamiltonian of these $n$ particles. The expansion coefficients
$b_{n}$'s are the so-called virial coefficients. The $n$th virial
coefficient $b_{n}$ is related to the solutions of up to $n$-body
problems. In this way, the complicated strongly correlated many-body
problem is converted to a set of relatively simple few-body problems.
For example, if we solve a two-body problem, we can obtain the second
virial coefficient $b_{2}=\left(Q_{2}-Q_{1}^{2}/2\right)/Q_{1}$ ,
and then we have the information about this many-body problem up to
the second order of the fugacity. The more we can solve the few-body
problems, the more accurate description of the many-body problem we
will acquire. All the other thermodynamic quantities can then be derived
from $\Omega$ via the standard thermodynamic relations, i.e., the
particle number $N=-\partial\Omega/\partial\mu$ , the entropy $S=-\partial\Omega/\partial T$
, and the energy $E=\Omega+TS+\mu N$ \cite{Fetter2003Q}.

All around this paper, we investigate the high-temperature thermodynamics
of a two-component (two hyperfine states, or, two spin states)\emph{
}harmonically trapped Fermi gas only up to the second order of the
fugacity. Generally, we are mostly interested in the contribution
arising from the interactions. Thus to this end, let us consider the
difference
\begin{equation}
\Delta b_{2}=b_{2}-b_{2}^{(1)}=\frac{Q_{2}-Q_{2}^{(1)}}{Q_{1}},\label{eq:3.2}
\end{equation}
 where the superscript ``$1$'' denotes the quantities of an ideal
gas. For a two-component Fermi gas, we easily find
\begin{equation}
\Delta b_{2}=\frac{1}{2}\sum_{\alpha}\left[e^{-E_{\alpha}^{rel}/\left(k_{B}T\right)}-e^{-E_{\alpha}^{(1),rel}/\left(k_{B}T\right)}\right],\label{eq:3.3}
\end{equation}
 where the summation runs over all possible states of the relative
motion of two fermions with different spin states. Therefore, the
spectrum of the relative motion determines the second virial coefficient.
The average energy per atom in unit of the Fermi energy $E_{F}=\left(3N\right)^{1/3}\hbar\omega$
($\omega$ is the trap frequency, and $N$ is the total atom number)
may be written as
\begin{multline}
\frac{E}{NE_{F}}=3\left(\frac{T}{T_{F}}\right)^{4}\left[3\int_{0}^{\infty}t^{2}\ln\left(1+ze^{-t}\right)dt\right.\\
\left.+6\Delta b_{2}z^{2}+2z^{2}\cdot T\frac{\partial\Delta b_{2}}{\partial T}\right],\label{eq:3.4}
\end{multline}
 where $T_{F}=E_{F}/k_{B}$ is the Fermi temperature. The fugacity
$z$ (or the chemical potential $\mu$) is determined by the total
particle number $N$ , and at a given temperature, can be solved from
\begin{equation}
\left(\frac{T}{T_{F}}\right)^{3}\left[z\int_{0}^{\infty}\frac{t^{2}e^{-t}}{1+ze^{-t}}dt+4\Delta b_{2}z^{2}\right]=\frac{1}{3}.\label{eq:3.5}
\end{equation}
 Subsequently, with the second virial coefficient $\Delta b_{2}$
and the fugacity $z$ in hand, the average energy per atom can easily
be obtained from Eq.(\ref{eq:3.4}). In addition, the other thermodynamic
quantites can also be acquired in the similar way.

\section{thermodynamics deeply inside the resonance width\label{sec:thermodynamics-deeply-inside}}

As we know, strongly interacting degenerate Fermi gases manifest universality
in the vicinity of the resonances \cite{Ho2004U,Hu2007U}. However,
for narrow Feshbach resonances, the appreciable effective range becomes
essential in this strongly interacting region. In this section, we
are going to study how the universality of a strongly interacting
Fermi gas is affected by the large effective range deeply inside the
resonance width.

In order to identify the effective range near the Feshbach resonance,
we may use the effective-range expansion of the scattering phase shift
$\phi_{0}$ in the low-energy limit \cite{Hazlett2012R,Ho2012A}.
We find the scattering length $a_{0}$ is expressed as
\begin{equation}
a_{0}=a_{bg}\left(1-\frac{\Delta}{B-B_{0}}\right),\label{eq:4.1}
\end{equation}
 and the effective range has a general form
\begin{equation}
r_{0}=-\frac{2\hbar^{2}}{m}\cdot\frac{\Delta}{a_{bg}\gamma\left(B-B_{0}-\Delta\right)^{2}},\label{eq:4.2}
\end{equation}
 where $a_{bg}$ is the background scattering length, $B_{0}$ is
the magnetic field strength at the resonance, $\Delta$ is the width
of the resonance, and $\gamma$ is the magnetic moment difference
between two atoms in the open channel and the molecule in the closed
channel. Deeply inside the resonance width, i.e., $B-B_{0}\ll\Delta$
, the $B$-dependence of the effective range is absent. The effective
range becomes a constant, i.e., $r_{0}\approx-2\hbar^{2}/\left(ma_{bg}\gamma\Delta\right)$
, and is determined only by the intrinsic properties of the resonance.

\begin{figure}
\includegraphics[width=1\columnwidth]{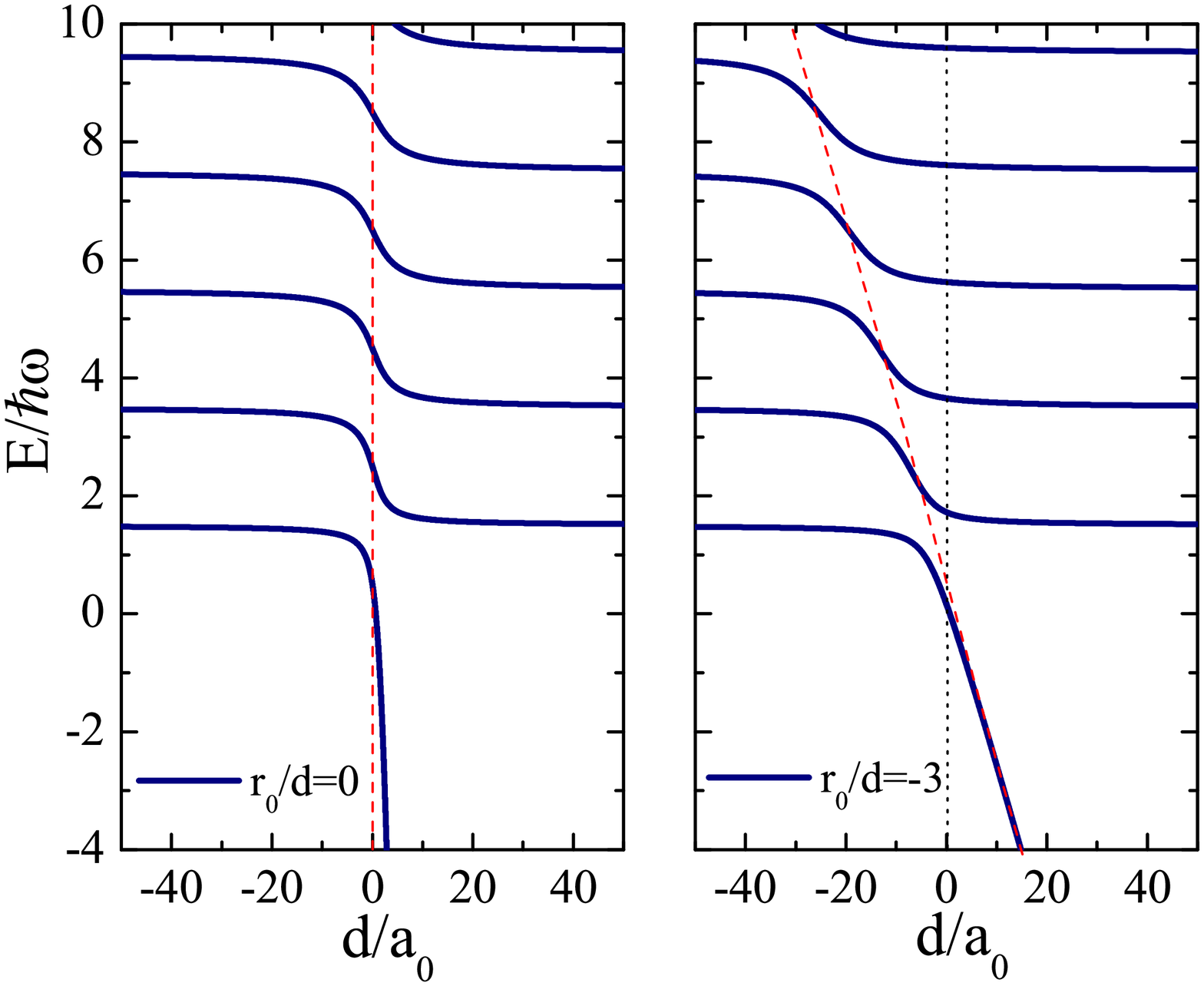}

\caption{(Color online) The relative-motion spectrums of two atoms in a harmonic
trap for broad (left) and narrow (right) resonances, respectively.
The red dashed lines are the eye guidance for the evolution between
two adjacent energy levels. }

\label{fig3}
\end{figure}

In this case, it is quite convenient to use the modified pseudopotential
introduced in \cite{Peng2011H,Peng2011N} to describe the two-body
interaction,
\begin{equation}
V_{0}\left(\mathbf{r}\right)=-\frac{4\pi\hbar^{2}}{m}\left(-\frac{1}{a_{0}}+\frac{1}{2}r_{0}k^{2}\right)^{-1}\delta\left(\mathbf{r}\right)\frac{\partial}{\partial r}r,\label{eq:4.3}
\end{equation}
 in which the \emph{constant} effective range is included besides
the scattering length. Here, $\mathbf{r}$ is the relative coordinate
of two atoms. After some straightforward and conventional algebra,
the relative-motion Schr\"{o}dinger equation of two atoms is reduced
to the following equation \cite{Peng2011H},
\begin{equation}
\frac{2\Gamma\left(-\frac{E}{2\hbar\omega}+\frac{3}{4}\right)}{\Gamma\left(-\frac{E}{2\hbar\omega}+\frac{1}{4}\right)}=\frac{d}{a_{0}}-\frac{r_{0}}{d}\frac{E}{\hbar\omega},\label{eq:4.4}
\end{equation}
 where $\Gamma\left(\cdot\right)$ is the Gamma function. We may easily
obtain the two-body spectrum of the relative motion from Eq.(\ref{eq:4.4})
as presented in Fig.\ref{fig3}. The effective range of the narrow
Feshbach resonance for $^{6}$Li at $B_{0}=543.3$G in the experiment
\cite{Hazlett2012R} is $r_{0}=-7\times10^{4}a_{B}$ ($a_{B}$ is
the Bohr's radius), and the trap frequency is around $\omega=2\pi\times1.3\sim3.6$kHz,
which determine the ratio of the effective range $r_{0}$ to the trap
characteristic length $d=\sqrt{2\hbar/\left(m\omega\right)}$ , i.e.,
$r_{0}/d\approx-2\sim-4$ . In our calculations, we choose $r_{0}/d=-3$
without loss of generality. As a comparison, we also plot the relative-motion
spectrum of two atoms across a broad resonance, where the effective
range is set to be $r_{0}/d=0$ in the broad-resonance limit. Obviously,
the properties of the narrow-resonance spectrum are quite different
from those of the broad resonances. For the narrow Feshbach resonance,
the energy-level transitions occur on the BCS side. Higher the energy
is, more deeply in BCS side the transitions are located. However,
all the energy-level transitions are located in the vicinity of the
broad resonance, and then such systems present the universality at
the unitarity limit.

\begin{figure}
\includegraphics[width=1\columnwidth]{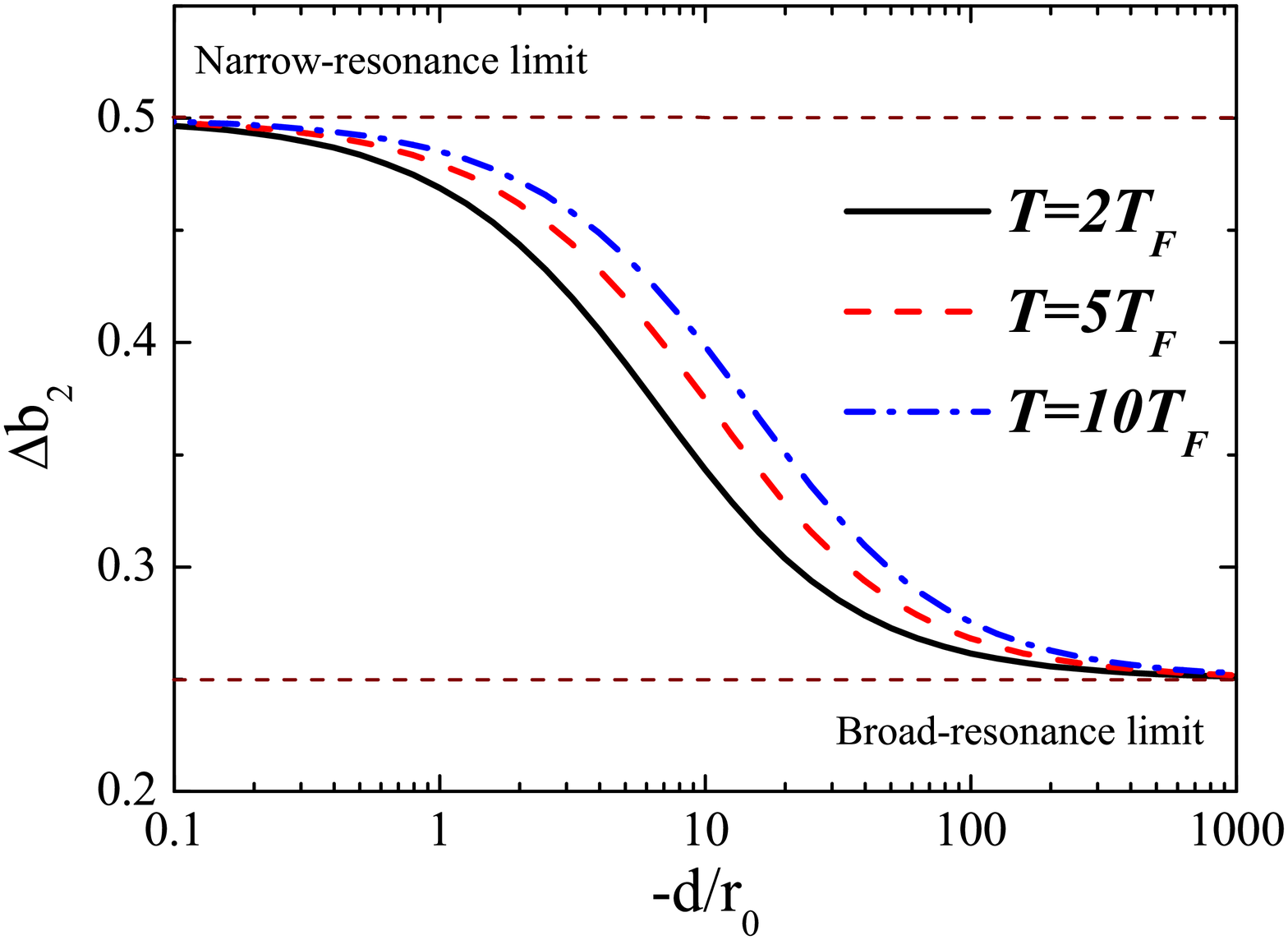}

\caption{(Color online) The second virial coefficients $\Delta b_{2}$ as functions
of the inverse effective range $-d/r_{0}$ at different temperatures.
Here, we have chosen the total atom number $N=1\times10^{5}$ . The
two horizontal dashed lines indicate the narrow- and broad-resonance
limits, respectively.}

\label{fig4}
\end{figure}

With the relative-motion spectrum of two atoms in hand, we can easily
obtain the second virial coefficient according to Eq.(\ref{eq:3.3}).
In order to demonstrate how the universality of a strongly interacting
Fermi gas is affected by the effective range, we plot in Fig.\ref{fig4}
the evolution of the second virial coefficient from the broad-resonance
limit to the narrow-resonance limit at different temperatures, as
the effective range increases. In the broad-resonance limit, it is
well-known that the second virial coefficient has a universal value
$1/4$ for a strongly interacting harmonically trapped Fermi gas \cite{Liu2009V}.
Interestingly, we find that this second virial coefficient will continuously
evolve to a larger universal value $1/2$ in the narrow-resonance
limit, which means the large effective range will result in a stronger
interaction between atoms at the resonance. Consequently, for a homogeneous
Fermi gas at the resonance, we find the second virial coefficient
$\Delta b_{2}^{hom}=2\sqrt{2}\Delta b_{2}=\sqrt{2}$ \cite{Liu2009V}.
This result is consistent with that of \cite{Ho2012A} (Note that
there is a $\sqrt{2}$ factor difference in the definition of the
second virial coefficient). 

\begin{figure}
\includegraphics[width=1\columnwidth]{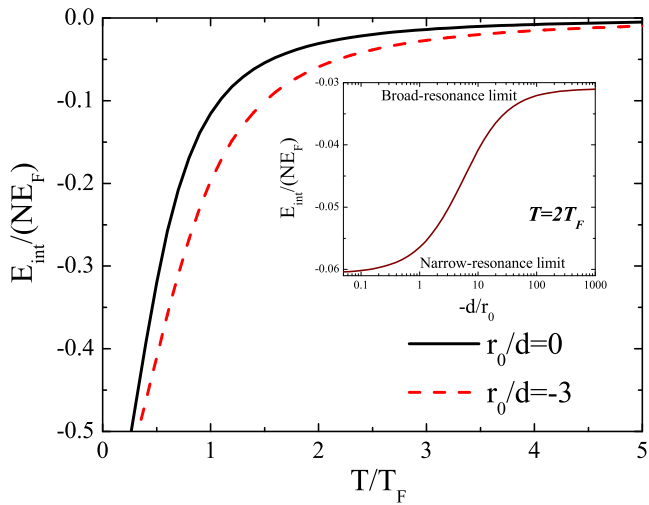}

\caption{(Color online) The temperature dependence of the interaction energy
of a strongly interacting harmonically trapped Fermi gas for a broad
resonance ($r_{0}/d=0$) and a narrow resonance ($r_{0}/d=-3$). The
insert is the evolution of the interaction energy with the effective
range from the broad-resonance limit to the narrow-resonance limit
at the temperature $T=2T_{F}$ . Here, we have chosen the atom number
$N=1\times10^{5}$ in our calculation.}

\label{fig5}
\end{figure}

According to Eq.(\ref{eq:3.4}), we can easily calculate the energy
of a strongly interacting harmonically trapped Fermi gas as well as
that of an ideal gas at a given temperature, and the difference gives
the interaction effect as shown in Fig.\ref{fig5}. At the resonance,
the interaction energy of a narrow resonance with appreciable effective
range is larger than that of the broad resonance. As the temperature
increases, the Fermi gas may approach to a classical ideal gas, and
the interaction effect vanishes. The evolution of the interaction
energy with the effective range from the broad-resonance limit to
the narrow-resonance limit is also illustrated in the insert of Fig.\ref{fig5}
at a specific temperature $T=2T_{F}$ . We find the Fermi gas becomes
more strongly interacted as the effective range increases at the resonance.

\section{thermodynamics far beyond the resonance width\label{sec:thermodynamics-far-beyond}}

Unlike the broad Feshbach resonance, it is relatively difficult to
experimentally explore the properties of a strongly interacting Fermi
gas quite deeply inside a narrow Feshbach resonance, due to the small
resonance width. In the narrow resonance experiment, it requires the
controllability of the magnetic field with high stability and accuracy
\cite{Hazlett2012R}. However, even outside the resonance width, the
narrow resonance system should also manifest unique properties which
are different from those of the broad resonance. In this section,
we are going to study the thermodynamics of a harmonically trapped
two-component Fermi gas across a narrow Feshbach resonance beyond
the resonance width.

As we already find in Eqs.(\ref{eq:4.1}) and (\ref{eq:4.2}), both
the scattering length and effective range are dependent on the magnetic
field strength $B$ across a narrow Feshbach resonance. In this case,
it is convenient to use the pseudopotential with an energy-dependent
scattering length $a_{0}\left(k\right)$ \cite{Bolda2002E},
\begin{equation}
V_{0}\left(\mathbf{r}\right)=\frac{4\pi\hbar^{2}a_{0}\left(k\right)}{m}\delta\left(\mathbf{r}\right)\frac{\partial}{\partial r}r,\label{eq:5.1}
\end{equation}
 to describe the two-body interaction, where the energy-dependent
scattering length $a_{0}\left(k\right)$ is defined from the scattering
phase shift $\phi_{0}\left(k\right)$ as \cite{Hazlett2012R,Ho2012A}
\begin{eqnarray}
a_{0}\left(k\right) & \equiv & -\frac{\tan\phi_{0}\left(k\right)}{k}\nonumber \\
 & = & a_{bg}\frac{\hbar^{2}k^{2}/m-\gamma\left(B-B_{0}-\Delta\right)}{\hbar^{2}k^{2}/m-\gamma\left(B-B_{0}\right)}.\label{eq:5.2}
\end{eqnarray}
 Then the relative-motion spectrum of two atoms is determined by the
following equation,
\begin{equation}
B-B_{0}=\frac{\left(E+\gamma\Delta\right)f\left(E\right)-E\cdot\frac{d}{a_{bg}}}{\left[f\left(E\right)-\frac{d}{a_{bg}}\right]\gamma},\label{eq:5.3}
\end{equation}
 where
\begin{equation}
f\left(E\right)=\frac{2\Gamma\left(-\frac{E}{2\hbar\omega}+\frac{3}{4}\right)}{\Gamma\left(-\frac{E}{2\hbar\omega}+\frac{1}{4}\right)},\label{eq:5.4}
\end{equation}
 and we have taken the approximation $k^{2}=mE/\hbar^{2}$ \cite{Peng2011H,Yip2008E,Suzuki2009T}.

\begin{figure}
\includegraphics[width=1\columnwidth]{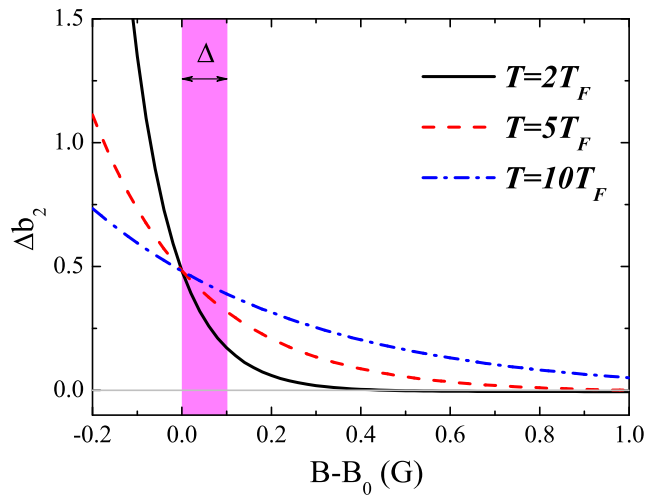}

\caption{(Color online) The second virial coefficient $\Delta b_{2}$ across
a narrow Feshbach resonance at different temperatures. Here, we consider
the narrow Feshbach resonance of a Fermi gas $^{6}$Li at $B_{0}=543.25$G
with the width $\Delta=0.1$G, background scattering length $a_{bg}=62a_{B}$
, and the magnetic moment difference $\gamma=2\mu_{B}$ ($\mu_{B}$
is the Bohr's magneton). We have chosen the trap frequency $\omega=2\pi\times2$kHz,
and the atom number $N=1\times10^{5}$ . The pink area is the conventional
strongly interacting region within the resonance width.}

\label{fig6}
\end{figure}

At a given magnetic field strength, we can easily solve the relative-motion
spectrum of two atoms, and subsequently evaluate the second virial
coefficient according to Eq.(\ref{eq:3.3}). In Fig.\ref{fig6}, we
plot the second virial coefficient in a real system of atoms $^{6}$Li
across a narrow Feshbach resonance at $B_{0}=543.25$G with the width
$\Delta B=0.1$G. At the resonance, i.e., $B=B_{0}$ , we find the
second virial coefficients at different temperatures almost cross
in one point quite near the universal value $1/2$ , which is consistent
with the result of the last section. In addition, we interestingly
find the second virial coefficient with nonzero value extends far
beyond the resonance width to deep BCS side, which means the Fermi
gas is still strongly interacted even far away from the resonance. 

\begin{figure}
\includegraphics[width=1\columnwidth]{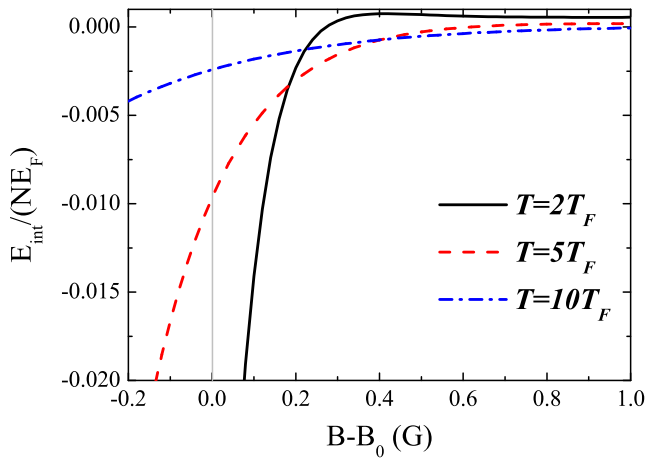}

\caption{(Color online) The interaction energy of a harmonically trapped Fermi
gas across a narrow Feshbach resonance at different temperatures.
We also consider the narrow Feshbach resonance of a Fermi gas $^{6}$Li
at $B_{0}=543.25$G with all parameters in Fig.\ref{fig6}.}

\label{fig7}
\end{figure}

The interaction energy of a harmonically trapped Fermi gas across
a narrow Feshbach resonance at different temperatures is presented
in Fig.\ref{fig7}. As we anticipate, there is indeed appreciable
interaction effect outside the resonance width. This is quite different
from the broad resonance case, in which we usually take the region
within the resonance width as the strongly interacting area. For narrow
resonances, due to large Fermi energy compared to the resonance width,
there may be still scattering resonance even above the zero-energy-resonance
point, i.e., $B=B_{0}$ \cite{Ho2012A,Gurarie2007R}. That is why
the Fermi gas is still strongly interacted even deep in the BCS side.

\section{about the energy-dependent interaction\label{sec:about-narrow-resonances}}

In our previous calculations, we just phenomenologically introduce
an energy-dependent interaction in the pseudopotential model to characterize
the narrow resonance. It is also very interesting to understand the
origin of such energy-dependent interaction. We will see that \emph{if
a metastable quasi bound state with a positive energy \cite{Landau1981Q}
is supported by the interatomic potential, it may result in a narrow
scattering resonance} \emph{with a large effective range}. 

\begin{figure}
\includegraphics[width=0.8\columnwidth]{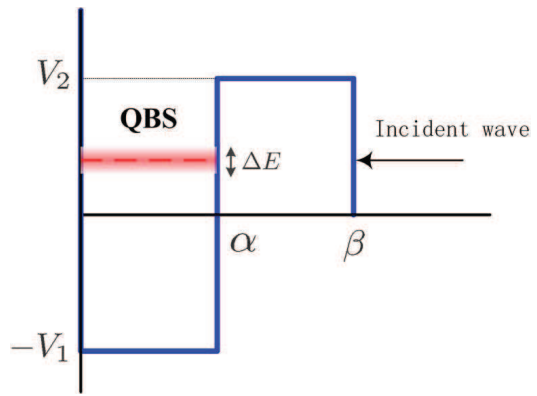}

\caption{(Color online) The square-well-barrier potential. Here, ``QBS''
is short for the quasi bound state.}

\label{fig1}
\end{figure}

To this end, let us consider a square-barrier-well potential, in which
the interaction potential between atoms takes the form,
\begin{equation}
V\left(r\right)=\begin{cases}
-V_{1}, & r<\alpha,\\
V_{2}, & \alpha<r<\beta\\
0, & r>\beta.
\end{cases},\label{eq:2.1}
\end{equation}
 as illustrated in Fig.\ref{fig1}. Although the real Feshbach resonance
is usually characterized by a two-channel model theory, the freedom
of the choice of the interaction potential between ultracold atoms
enables us to extract the key information from this simple single-channel
model. This toy model is effectively equivalent to the two-channel
model, and is already sufficient to realistically mimic the narrow
Feshbach resonance \cite{Schneider2013T}. 

\begin{figure}
\includegraphics[width=1\columnwidth]{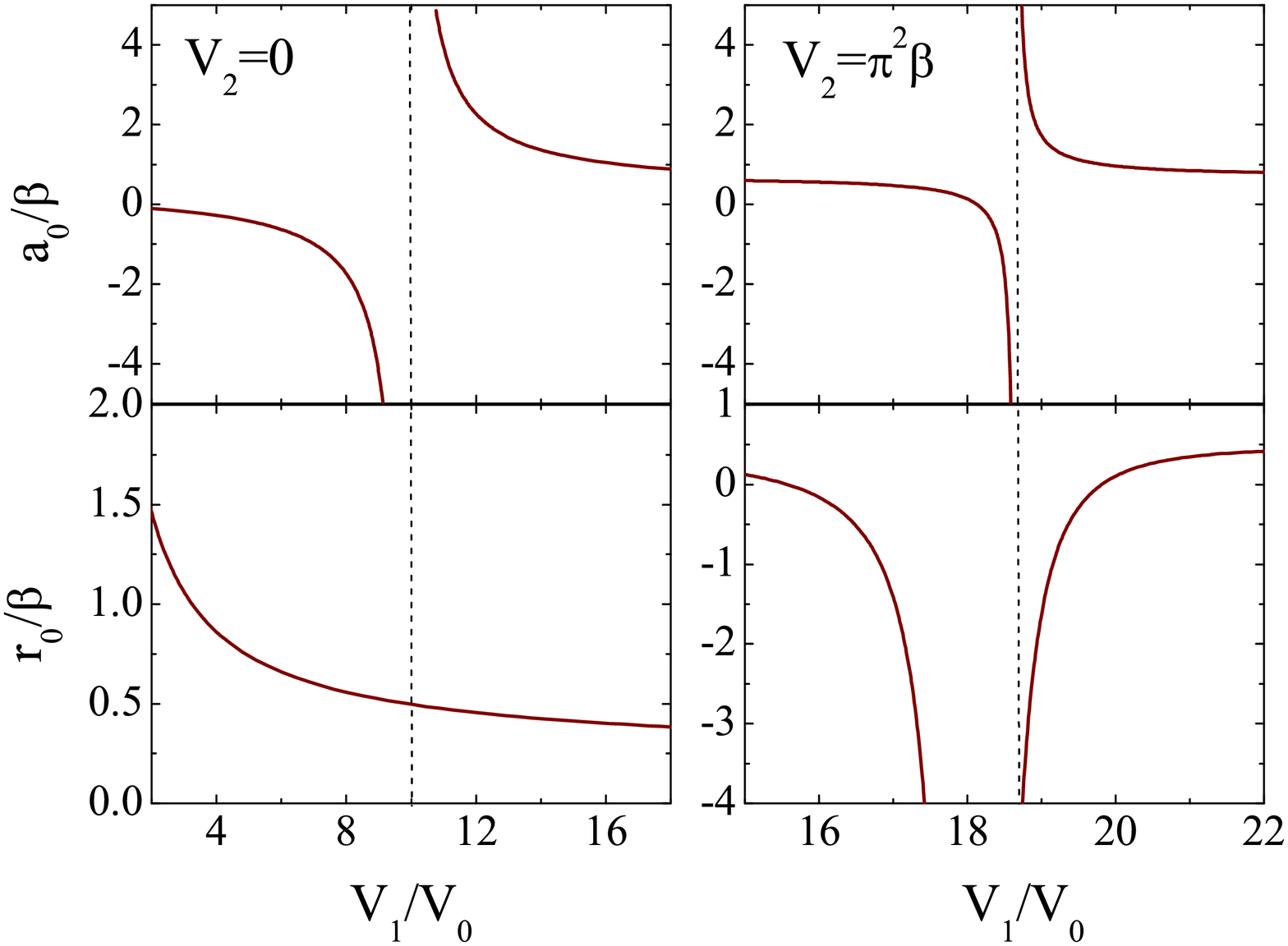}

\caption{(Color Online) The scattering length $a_{0}/\beta$ and the effective
range $r_{0}/\beta$ as functions of the square-well depth $V_{1}/V_{0}$
in absence (Left panel) and in presence (Right panel) of a barrier,
where $V_{0}=\sqrt{\hbar^{2}/m\beta^{2}}$ and $m$ is the atomic
mass. Here, we have chosen $\alpha/\beta=0.5$ .}

\label{fig2}
\end{figure}

We may easily solve the scattering problem within this interatomic
potential, in presence and absence of the barrier, respectively. The
scattering length $a_{0}$ and effective range $r_{0}$ can be evaluated
from the scattering phase shift $\phi_{0}$ ,
\begin{equation}
k\cot\phi_{0}=-\frac{1}{a_{0}}+\frac{1}{2}r_{0}k^{2}.\label{eq:2.2}
\end{equation}
From Fig.\ref{fig2}, we find the presence of the barrier raises a
much larger effective range near the resonance compared to the case
without the barrier, which means this square-well-barrier model indeed
causes an energy-dependent interaction. This is because,\emph{ in
presence of the barrier, a quasi bound state with a positive energy
may have a finite life time, and results in a relatively small energy
broadening} \emph{due to the uncertainty principle},\emph{ compared
to the case without the barrier}. In this case, when the incident
atoms energetically approach this bound state, a narrow scattering
resonance occurs. Similarly, in the real two-channel Feshbach resonance,
the lifetime of the bound state in the closed channel is prolonged
due to the large Fermi energy of atoms in the open channel \cite{Diener2004,Gurarie2007R},
and thus induces a relatively narrow Feshbach resonance.

\section{conclusions\label{sec:conclusions}}

In summary, we have presented the virial expansion in studying the
high-temperature thermodynamics of a \emph{harmonically trapped} Fermi
gas across a narrow Feshbach resonance. Deeply inside the width of
a narrow resonance, the second virial coefficient is obtained based
on the two-body spectrum in a harmonic trap. We find the second virial
coefficient undergoes a continuous evolution from $1/4$ in the broad-resonance
limit to $1/2$ in the narrow-resonance limit, as the effective range
increases. This results in a larger interaction energy for a narrow
resonance compared to that of a broad resonance. Besides, contrary
to our knowledge of a broad resonance, there is still considerable
interaction effect far beyond the width of a narrow resonance. Finally,
we present a simple toy model for helping us understand the origin
of the energy-dependent interaction in narrow Feshbach resonances.
All our results can be directly tested in current narrow Feshbach
resonance experiments, which are generally carried out in a harmonic
trap.
\begin{acknowledgments}
We are appreciated for fruitful discussions with Shina Tan. This work
is supported by NSFC (No. 11004224 , No. 11204355 and No. 91336106),
NBRP-China (No. 2011CB921601), CPSF (No. 2012M510187 and No. 2013T60762)
and programs in Hubei province (No. 2013010501010124 and No. 2013CFA056).\end{acknowledgments}

\end{document}